# UAV Swarm Path Planning with Reinforcement Learning for Field prospecting

**Alejandro Puente-Castro**\*
Faculty of Computer Science, CITIC
University of A Coruna
A Coruna, 15007, Spain

**Daniel Rivero**
Faculty of Computer Science, CITIC
University of A Coruna
A Coruna, 15007, Spain

**Alejandro Pazos**
Faculty of Computer Science, CITIC
University of A Coruna
A Coruna, 15007, Spain
Biomedical Research Institute of A Coruña (INIBIC)
University Hospital Complex of A Coruna (CHUAC)
A Coruna, 15006, Spain

**Enrique Fernandez-Blanco**
Faculty of Computer Science, CITIC
University of A Coruna
A Coruna, 15007, Spain

June 3, 2021

## Abstract

Unmanned Aerial Vehicle (UAV) swarms adoption shows a steady growth among operators due to the benefits in time and cost arisen from their use. However, this kind of system faces an important problem which is the calculation of many optimal paths for each UAV. Solving this problem would allow a to control many UAVs without human intervention at the same time while saving battery between recharges and performing several tasks simultaneously. The main aim is to develop a system capable of calculating the optimal flight path for a UAV swarm. The aim of these paths is to achieve full coverage of a flight area for tasks such as field prospection. All this, regardless of the size of maps and the number of UAVs in the swarm. It is not necessary to establish targets or any other previous knowledge other than the given map. Experiments have been conducted to determine whether it is optimal to establish a single control for all UAVs in the swarm or a control for each UAV. The results show that it is better to use one control for all UAVs because of the shorter flight time. In addition, the flight time is greatly affected by the size of the map. The results give starting points for future research such as finding the optimal map size for each situation.

***K*eywords** UAV swarm · Path planning · Reinforcement Learning · Q-Learning · Artificial Neural Network · Agriculture

## 1 Introduction

New applications of Unmanned Aerial Vehicles (UAV or drones) swarms are developed nearly everyday for different problems, such as crop monitoring [1, 2], forestry activities [3], space exploration [4, 5], or military and rescue missions [6]. Main reason for that popularity lays on the advantages of the UAVs, such as low cost, great maneuverability, safety, and convenient size for certain kinds of maneuvers [7]. However, they also has disadvantages being the main one the battery consumption, which limits the flight time. When used in a group or swarm their flight time limitations are reduced. In other words, several UAVs flying simultaneously allows many tasks to be carried out in less time.

The use of UAV swarms also allows for fault tolerance. If any UAV in the group is unable to continue operating, the other UAVs can replace it and thus be able to complete the task as well.

---

\*Corresponding author: a.puentec@udc.es

When the first flight tests were carried out, they needed as many operators as UAVs, which increases the cost of operation considerably. More recently, advances have been registered in the creation of algorithms [8] and telecommunications [9] necessary for the control of the entire swarm with only one user capable of executing the systems. These advances grant better and faster communications between UAVs and grant the fast calculation of collision avoidance paths, so that less human intervention is required if there is any risk. Thus, the operation is less expensive because it requires fewer personnel.

To deal with the complexity of this kind of development, in Swarm Intelligence, different algorithms are proposed capable of coordinate numerous agents simultaneously. This coordination is based on a group of individuals that follows common simple rules in a self-organized and robust way [10].

Today, most of these path planning algorithms are of military application. The few civilian applications are usually to follow or reach targets, such as mapping paths through cities. There are few systems oriented to agricultural and forestry use, specially dedicated to the optimization of the field prospecting tasks.

This objective of this paper is to develop a system for solving the Path Planning problem in 2D grid-based maps with different number of UAVs using Q-Learning techniques. The main contributions of this paper can be listed as: first, a system capable of calculating the optimal flight path for the UAVs in a swarm for field coverage in prospecting tasks; second, a system capable of calculating the flight path of any number of UAVs and with different map sizes; third, a system capable of calculating paths without the need to establish targets or provide information other than the actual state of the map; and fourth, a study on the difference in the results of using a global ANN for all UAVs and using an ANN per UAV.

The outline of this paper follows this structure: in Section 2, there is a brief summary of the current state-of-the-art; in Section 3, an explanation is given of the technical aspects necessary for the development of the proposed algorithm; in Section 4, there is a summary of the results obtained from the experimentation process; in Section 5, the results obtained are discussed; in Section 6, the conclusions obtained after reviewing the obtained results are listed; finally, in Section 7, the possible works and studies in which they can derive the problem to be addressed are listed.

## 2 Background

In the state of the art, there are several approaches, two of which are particularly noteworthy: the first one makes use of Reinforcement Learning (RL) [11]; while the second one points its attention out to Evolutionary Computing (EC) [12].

RL algorithms for path planning are the most abundant in the state of the art. For example, Xie et al. use the Q-Learning strategy for three-dimensional path planning [13]. They introduced the concept of Heuristic Q-Learning. This allows a better adjustment of the reward based on the current state and the available actions for speeding up the convergence process to the optimal result. Roudneshin et al. use Deep Q-Learning to control swarms composed of UAVs and heterogeneous robots [14]. Instead of a pure UAV approach, this work adds heterogeneous terrestrial robots to the swarms. However, this is a problem of swarm path planning with greater difficulty than using purely UAVs. This increase in the difficulty of the problem is due to the different limitations presented by air and land vehicles. Thus, a land vehicle can encounter non-geographic obstacles and has more limited movements.

Others, however, make use of the RL algorithm called SARSA [15] strategy as Luo et al. where they tested their Deep-SARSA algorithm in dynamic environments, where the obstacles might change [16]. They show the behavior of their system in changing environments, which reinforces their usefulness in the real world. The model requires a pretraining phase, which could limit its deployment in novel environments due to the time required for pretraining.

In [17], Speck et al. combine object-focused learning with this algorithm in a very efficient decentralized approach when generalizing. This capacity for generalization might be limited because it was developed for fixed-wing UAVs. The configuration of these UAVs limits the range of application of the system to cases where it is optimal to use fixed-wing UAVs because these aircraft do not have stationary flight capabilities.

On the other hand, there are EC-based methods. For example, Duan et al. combine a genetic algorithm with the VND search algorithm [18]. An initial individual is generated based on the heuristics of its nearest neighbors and the rest of the initial individuals are configured as random. The use of the closest neighbors limits the generation of individuals. Especially in the case of many equally close neighbors. In that situation, it is necessary to establish a criterion to determine whether the individual is a member of a group. Recently, Liu et al. employ Genetic Algorithms to adjust ANN for flight path generation [19]. Relying only on the ANN for path computation makes their system dependent on more parameters than weights. Therefore, other parameters such as learning rates or adjusting the architecture of the ANN should be adjusted.

There are other methods applied to path planning with UAV swarms. For example, Vijayakumari et al. make use of Particle Swarm Optimization for optimal control of multiple UAVs in a decentralized way [20]. They manage to simplify the computation of the problem by means of discretization. They rely on distances for collision avoidance. Although this is a dynamic variable, in certain types of non-stationary flight UAVs, such as fixed-wing UAVs, it does not guarantee collision avoidance. In these cases, a metric that predicts the state of the UAV and the obstacle in future instants is of interest and thus makes a decision. Otherwise, the UAV would continue to move forward while the decision is being computed. Li et al. make use of Graph Neural Network for path computation in robotic systems. Thus, they achieve more capacity for generalization in the face of new cases than other more widely used techniques [21]. Since they are dealing with two ANNs, previous training is necessary in different and very varied cases. Otherwise, ANNs might be overfitted in several flight areas and swarm structures.

As it was shown, in the associated literature, systems often require extra map information such as targets or distance maps. In addition, they use maps with fixed numbers of cells. The aim of this work is to propose a system without the need for extra map information and that works with any map size.

## 3 Materials and Methods

For the complete development of the paper, four main factors have been taken into account: first, the flight environment; second, the chosen method; third, the estimation of the energy consumption of the UAV; and finally, the necessary metrics to know the goodness of the developed models.

### 3.1 Flight Environments

Previously published works, described in Section 2, used fixed squared maps of dimensions between $10 \times 10$ and $20 \times 20$. The approach presented in this work takes a wider point of view allowing the use of arbitrary polygons as maps, e.g the one presented in Figure 1(a). To do this, the following steps have been followed:

1. The minimum bounding rectangle (MBR) of the map has to be calculated such as in Figure 1(b). The map polygon is surrounded by a rectangle of the smallest possible size based on the combined spatial extent of one or more selected map features [22]. In this case, based on its vertices.
2. The resulting MBR is is divided into cells like in Figure 1(c). Cells in the resulting grid have to be labelled as visitable and non-visitable.

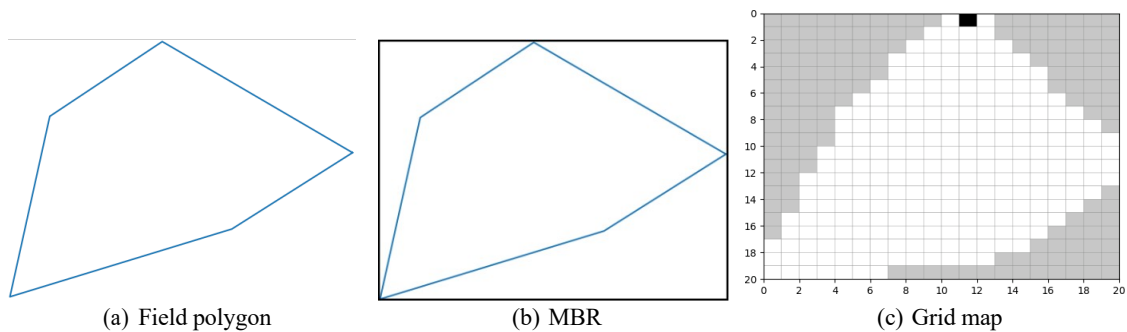

(a) Field polygon    (b) MBR    (c) Grid map

Figure 1: (a): Example of the polygon representing the area of the indicated field. (b): The same polygon (blue) surrounded by its minimum bounding rectangle (black). The MBR must be as close as possible to the polygon. (c): The MBR divided into cells. The black cell is the starting point of the UAVs. The gray cells are those that cannot be flown over and the white cells are those that can be flown over.

### 3.2 Proposed Method

#### 3.2.1 Reinforcement Learning

Reinforcement Learning (RL) [11] was chosen as the technique to calculate the optimal path to cover the maps by the UAVs. In this technique, the agents learn the desired behaviour based on a trial-and-error schema of tests executed on an interactive and dynamic environment [11, 23, 24]. The goal is to optimize the behavior of the agent in respect to a reward signal that is provided by the environment. The actions of the agent can also affect the environment, complicating the search for the optimal behavior [25].

All RL algorithms follow a common structure, the only difference is the learning strategy. There are several types of these strategies which allow the models to deal with different problems. For this paper, it has been decided to use a variant known as Q-Learning. The main motivation is that, unlike other variants, it does not require a model of the environment.

### 3.2.2 Q-Learning

Classic Q-Learning algorithms (Algorithm 1) are a kind of off-policy RL algorithms, so the agents can use their experience to learn the values of all the policies in parallel, even when they can follow only one policy at a time [26]. It follows a model-free strategy [27], where the agent gets knowledge by following a policy only by trial-and-error. In this way, Q-Learning converges towards the optimal solution in a greedy way. Allowing the optimal solution to be reached without being dependent on the decision-making policy. In other words, it makes decisions based purely on the environment surrounding the agent and its interactions with it. In this way, it is guaranteed that the system can work with different types of environments without having to search for the optimal policy that works in all of them. The "Q" in Q-learning stands for quality, which tries to represent how useful a given action is in gaining some future reward.

---

**Algorithm 1** $Q$-learning: Learn function $Q : S \times A \to R$

**Require:**
  States $S = \{1, \ldots, n_x\}$
  Actions $A = \{1, \ldots, n_a\}$,  $A : S \Rightarrow A$
  Reward function $R : S \times A \to R$
  Black-box (probabilistic) transition function $T : S \times A \to S$
  Learning rate $\alpha \in [0, 1]$, typically $\alpha = 0.1$
  Discounting factor $\gamma \in [0, 1]$
  **procedure** Q-LEARNING($S, A, R, T, \alpha, \gamma$)
    Initialize $Q : S \times A \to R$ arbitrarily
    **while** $Q$ is not converged **do**
      Start in state $s \in S$
      **while** $s$ is not terminal **do**
        Calculate $\pi$ according to Q and exploration strategy ($\epsilon$ greedy)
        $a \leftarrow \pi(s)$
        $r' \leftarrow R(s, a)$ ▷ Receive the reward
        $s' \leftarrow T(s, a)$ ▷ Receive the new state
        $Q(s, a) \leftarrow r + \gamma \times \arg\max_{a'}(Q(s', a'))$ ▷ Bellman Equation
        $s \leftarrow s'$
    **return** $Q$

---

The main difference between these algorithms and other RL algorithms is that they determine the best action based on the values in a table. The table is known as Q-table and the values as Q-values. These values determine how rewarding it would be to perform each action given the current state of the environment. From these values, the action with the highest value for each state is chosen. Typically, models are trained by combining their previous predictions with Bellman's equation (Eq. 1). The equation has different elements: $Q(s, a)$ is the function that calculates the Q-value for the current state ($s$), of the set of states $S$, and for the giving action ($a$), of the set of actions $A$, $r$ is the reward of the taken action in that state and it is computed by the reward function $R(s, a)$, $\gamma$, is the discount factor and $\arg\max_{a'}(Q(s', a'))$ is the maximum computed Q-value of the pair ($s'$, $a'$). The pair ($s'$, $a'$) is a potential next state-action pair. ($s'$ is the next state and it is given by the transition function $T(s, a)$ which returns the state resulting from performing the selected action. The $a'$, is each one of the available actions. Through an initial exploration process, the chosen value for $\gamma$ is 0.91.

$$Q(s, a) \leftarrow r + \gamma \times \arg\max_{a'}(Q(s', a')) \quad (1)$$

Alternatively, in the last years, a modification has to arise called Deep Q-learning. These methods differ from the classic Q-Learning [28] in that it seeks to improve the calculation of the Q table through Machine Learning [29] or Deep

Learning models [30]. The model is able to abstract enough knowledge to infer the values of the Q table. In this way, it is possible to overcome Bellman Equation's bias issues in some scenarios [31].

The aim is to improve classical Q-Learning by using small ANNs. In this work, authors chose to use densely connected ANNs with two layers. Using only two layers learning and decision-making are usually done in less time compared with convolutional deep ANNs [32] that other author propose in their papers. For this purpose, the following steps are followed in each Q-Learning experiment:

1. Build the ANN model o models based on the chosen configuration.
2. Employ the model or the models to determine Q-table values in order to choose the best action for each UAV in the swarm.
3. Train the model or the models based on the consequences of taking each of the selected actions.
4. Select those cases where the flight time required to explore the entire map is lower.

Through empirical experimentation, a network formed by two dense layers [33] has been chosen: the first one with 167 neurons and linear activation function [34] and the second one with 4 neurons and softmax activation function. The chosen optimizer for the ANN was RMSprop [35]. Maps are the only input of the network (Figure 2). Thus, ANN does not need more information than what is included in the maps.

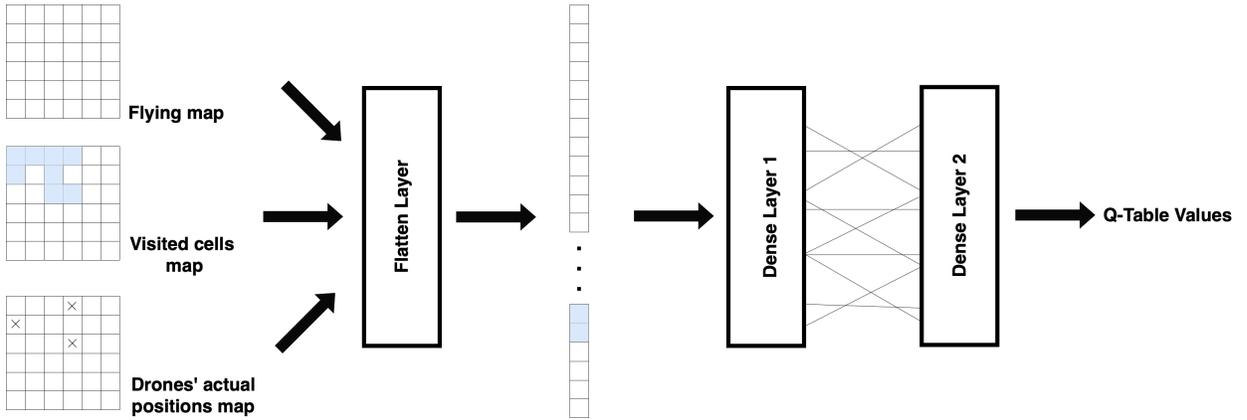

Figure 2: Diagram showing how data is processed in the ANN in order to get Q-values of the Q-table for a given state of the environment. The maps are combined into a multidimensional matrix and then flattened into vectors. These vectors are used to abstract the knowledge for computing Q-values.

From this point, the system could be used in two different approaches with no clear advantage for any of them. First, a single ANN is developed and used as the control for each of the UAVs. Therefore, all UAVs are going to have the same architecture and weights and their behavior depends on the current state of the UAV. On the other hand, each UAV can have a different ANN therefore its answer would not be only the result of the state but also the weights and architecture codified on them.

In all Q-Learning problems, part of the actions is taken randomly with probability epsilon ($\epsilon$ = 0.47), and with probability $1 - \epsilon$ the action with the highest Q-value for that state is taken. The sequence of actions taken by an agent for a given $\epsilon$ until it reaches an end condition (task completed, end of time...) is known as an episode. In each episode, the task is restarted from the beginning. As episodes occur during the experimentation, the $\epsilon$ value is reduced multiplying it by a reduction factor equal to 0.93. In this way, the choice of actions falls more on the calculated Q-values and less by random selection. To avoid overfitting, $\epsilon$ is prevented from reaching a value very close to 0 by setting the minimum value 0.05. Both values were chosen through a previous exploratory study.

### 3.2.3 Rewards

In order to prioritize the UAV to move to unvisited areas, the reward must be the highest of all (Table 1. In addition, it is important that it increases as fewer cells are left undiscovered (Eq. 2). This is known as Hill-Climbing [36]. Another reward is required for cells that have already been visited. Thus, the UAV has a reward in case it is more optimal to fly over an already visited cell to reach an unvisited one than to go around it (for example, when there are spurious cells left unvisited). To prevent UAVs from flying into cells that they cannot visit, they are given the lowest. The choice of the selected reward values was made through an initial exploratory process.

|  | **Reward** |
|---|---|
| new cell base reward | 358.74 |
| visited cell reward | -31.14 |
| non-visitable cell | -225.17 |

Table 1: Table with the assigned rewards to each kind of cell each UAV visits. The initial values chosen for the rewards by means of a previous random exploration where the best combinations of rewards have been selected.

$$\text{new cell reward} = \text{new cell base reward} \times (1 + \frac{max(\text{rows, columns})}{\text{visited cells}}) \qquad (2)$$

### 3.2.4 Flying Actions

The possible movements or actions ($a$) from the set of actions $A$ that UAVs can take were codified. Thus, all possible movements are encoded to a discrete list of values.

Despite the natural complexity of UAV flight, the possible movements have been simplified into straight movements. This allows having more easily interpretable flight paths in a map divided into cells. If not, it could be the case that a UAV draws a curve passing over the corner of a cell without passing through the cell entirely. In that case, there is the dilemma of having to mark that cell as visited or not (Figure 3). Not tracing curves ensures that the graphic data obtained with UAVs always have the same angle and are easier to combine.

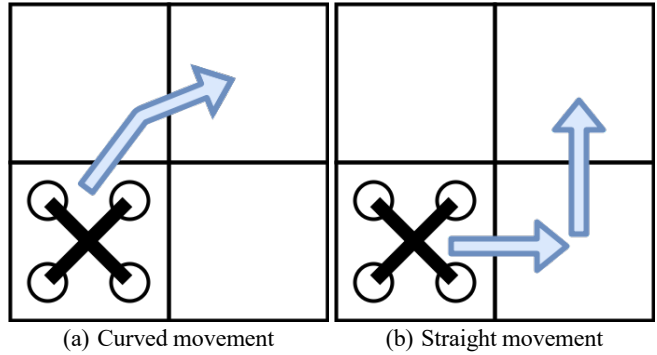

(a) Curved movement     (b) Straight movement

Figure 3: Comparison of curved movements with straight movements Curved motions produces more easily interpreted flight paths. Moreover, they are limited to the atomicity of the map cells.

### 3.2.5 Memory Replay

In most of the State of the Art, the experience obtained by agents from the environment is reinforced with the Memory Replay technique. Memory Replay is a technique were the model is trained with a set stored observations called memory. The observations contain a variety of information. like taken actions and their reward. It improves sample efficiency by repeatedly reusing experiences and helps to stabilize the training of the model [37]. It is important that the memory contains as many recent observations as possible, but it has a maximum size in order to optimize computational resources. For this reason, the memory follows a First-In-First-Out scheme for eliminating old observations.

Each UAV in the group has its own memory. In its memory, it stores observations with the actions that the UAV itself takes. At no time the actions of other UAVs are stored. This avoids adding noise to the information. The fact that an action is not correct for one UAV does not imply that it is correct for the others since they can be in different positions on the map.

The size of the memory can greatly vary the final results [38]. For this study, a memory size of 60 actions with their respective rewards was chosen after an exploration process. It is important to have a large value with respect to the number of map cells because in the first iterations of the process UAVs make many errors. Thus, learning from most of the errors and training the model multiple times with them will help to avoid them and, thus, achieve a more efficient solution.

### 3.2.6 Workflow

Schema shown in Figure 4 summarizes the main workflow of the proposed method. Starting from the read of the initial data, that is the vertices of the area to be covered and the initial positions of the drones, the maps are reconstructed. After that, by using those maps, where the ANN is trained with the Q-Learning technique [28]. In a part of the experiments it is a global ANN and in another part an ANN per UAV. This is going to determine the best action for each UAV.

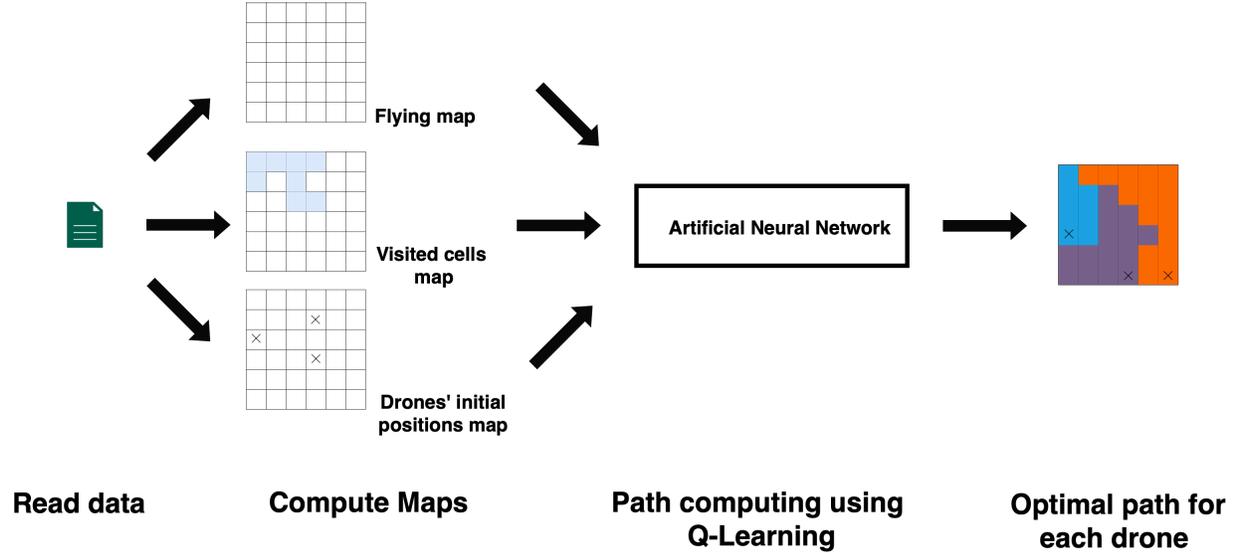

Figure 4: The workflow diagram of the study. The map data and the position of the UAVs are read from a file. From the data read, the maps that the system will use are constructed. Using the Q-Learning technique, the best possible path is calculated for each UAV so that the task is completed.

## 3.3 Battery Estimation

As in the works discussed in Section 2, the authors have not found a standardized method to predict battery power consumption during flight. This is because the consumption depends on the UAV configuration and flight conditions. Normally, most commercial UAVs send to their mobile apps the amount of energy they have leftover from time to time. On the other hand, there are more and more websites that help calculate how much battery time is left. The lack of standardization is due to the influence of many variables. The incident wind, the number of direction changes, speed, and many other variables greatly affect the time of flight.

As a solution to the problem of battery consumption, the swarm is forced to find a solution in the remaining battery time that is minimum among the UAVs of the swarm. In this way, it is expected that in a limited time the UAVs will try to get as close as possible to the desired solution. In this study, it has been chosen to assume that the UAVs all have a maximum energy load that allows them to fly for 30 minutes because no realistic calculation of the remaining battery time discount has been found.

## 3.4 Performance Measures

As performance measures, the most common ones will be taken into account: the time needed to find the solution, the percentage of correct actions out of all actions taken, and the evolution of the map coverage.

It is necessary to find a system that finds the solution to the problem as quickly as possible. Thus, it will require less operator time and battery consumption when used in real fields. Low battery consumption indicates that the paths are as short as possible. In addition, due to the charging time of the UAV batteries, low battery consumption might allow the user to do more work without having to stop charging the batteries. This makes it the measure of greatest interest and the one mainly chosen. For this purpose, we will look for the episodes with the shortest execution time ($ET$), which is computed as the difference between the actual time when the episode finished or $TE_1$ and the actual time when the episode started or $TE_0$ (Eq. 3). It is important to compute this coverage for each taken action in each individual episode in order to get the curve that relates the evolution of the number of cells discovered by the agents with the number of total actions that they carry out. The greater the growth of the curve means that fewer movements are needed to reach the solution. This implies that the paths they take have fewer cycles and are therefore more efficient. In Figure 5 there is an example plot of the curve for one ANN per UAV using two UAVs. In its evolution compared to using a the same ANN for all UAVs (a global ANN), growth is higher, and it reaches total coverage much faster.

$$ET = TE_1 - TE_0 \tag{3}$$

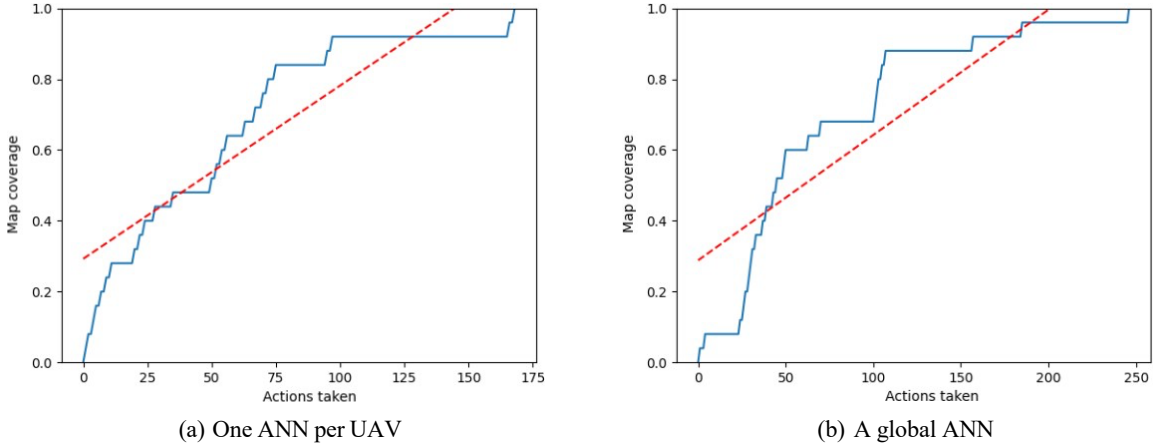

(a) One ANN per UAV

(b) A global ANN

Figure 5: Example of curves of the evolution of map coverage (y-axis) as a function of the number of actions taken by two UAVs (x-axis). The dashed line represents the coverage growth trend. Comparing the case of one ANN per UAV with its respective with a global ANN it can be seen that the case with a global ANN takes more actions for flying the whole map.

Even though time is used as a measure of performance, it is also needed to know the length of the paths UAVs take for each episode. Depending on the configuration of the UAV and the way the UAV flies, the battery consumption can greatly vary from one to another. Therefore, it is interesting that the path is as short as possible. It can be understood as the largest possible number of valid actions to be taken. Valid actions have been defined as those where a new cell is discovered. That is, without loops or passing through cells that cannot be visited. Therefore, it is of interest to know what fraction (*PA*) of valid actions (*VA*) out of all actions taken by the UAVs (*TA*) (Eq. 4). The closer to 1, the better.

$$PA = \frac{VA}{TA} \tag{4}$$

Knowing how the total map coverage evolves makes it possible to distinguish which methods are better. It is calculated as the fraction of cells that have been visited divided by the total number of cells. Sometimes, the operational resources available (number of UAVs, battery levels...) might not be sufficient to overfly the selected terrain in its entirety. Even so, in such cases, it is important to cover as large an area as possible. That is, a system in which it is able to get closer and closer to 100% map coverage is ideal. The closer to 1, the better.

## 4 Results

A set of combinations of map sizes and number of UAVs has been defined for conducting the experiments and subsequent analysis of the results. For the analysis of the results obtained, factors such as the evolution of the time required to explore the map and the percentage of actions performed by each UAV have been taken into account.

### 4.1 Experiment Design

To test the capabilities of the system proposed in this paper, 25 experiments have been designed. In each of them, the configuration of the ANNs, the number of UAVs, and the size of the map are changed.

| Approach | Map size | Number of UAVs |
|---|---|---|
| Baseline | 5×5 | 1 UAV |
| | 6×6 | 1 UAV |
| | 7×7 | 1 UAV |
| | 8×8 | 1 UAV |
| | 9×9 | 1 UAV |
| One ANN per UAV | 5×5 | 2 UAVs |
| | | 3 UAVs |
| | 6×6 | 2 UAVs |
| | | 3 UAVs |
| | 7×7 | 2 UAVs |
| | | 3 UAVs |
| | 8×8 | 2 UAVs |
| | | 3 UAVs |
| | 9×9 | 2 UAVs |
| | | 3 UAVs |
| Global ANN | 5×5 | 2 UAVs |
| | | 3 UAVs |
| | 6×6 | 2 UAVs |
| | | 3 UAVs |
| | 7×7 | 2 UAVs |
| | | 3 UAVs |
| | 8×8 | 2 UAVs |
| | | 3 UAVs |
| | 9×9 | 2 UAVs |
| | | 3 UAVs |

Table 2: Table with the 25 experiments performed. Each one of them with different configuration. The experiments for an ANN per UAV for a single UAV have been omitted because it is the same as using a global network for a single UAV.

The experiments were carried out in a square cell map as in those cited in Section 2.

The aim og this experiments is to identify the best controller for the UAVs. Two are the approaches at this point, one ANN per UAV and one ANN for all UAVs. Both approaches were confronted in the same maps and with the same UAVS. The results that can be seen in Table 2. As can be seen in the table, the experiments with an ANN for a UAV have been omitted when there is only a single UAV. Using one ANN for only one UAV would be the same as using a global ANN for only one UAVS. Therefore, it has been simplified to execute only once with a global ANN for obe UAV and it is referred to as baseline (Figure 6). Thus, it is taken as the starting point of the experimentation taking it as the simplest case, which is to control a single UAV.

Since it is important that the system is able to operate with any number of UAVs, each selected map type was tested with an increasing number of UAVs. To be more precise, separate experiments have been performed with 1, 2, and 3 UAVs. Thus, it is proved that the system is able to adapt to a different number of UAVs.

As in the papers mentioned in Section 2, all the maps chosen are grid maps. The experiments were performed with 5×5, 6×6, 7×7, 8×8 and 9×9 cell grid maps. Having different map sizes provides insight into the capabilities of the system in the face of unfixed map sizes.

In addition, the number of cells in the chosen flight environment is smaller than other cited papers. The cost of flying over large maps is a major constraint. By making one stop per cell to photograph the surface of the map each cell contains means that in very large maps the UAVs have to make many stops considerably affecting their battery. Dividing the map into fewer cells reduces the number of stops and starts made by each UAV decreasing their energy consumption.

Another factor to consider is the area of land that each cell represents. The larger, the better, the more information each image contains and the more favorable it is for further processing. These cells must contain an adequate surface area size for each type of activity performed. For example, in tasks such as water stress [39], in which one flies at a height of 12 meters, the area size of the map contained in each cell is enormous.

In many countries the distance from the position to which it flies is limited to what a UAV can fly. For example, in many European countries it is 500 meters or additional measures would have to be taken that not all operators can overcome [40]. Therefore, as it is known that the terrain cannot be too large for the system to be used in a generic way, it is not necessary to use maps with numerous cells. For example, 400-cell maps such as those used in some papers described in Section 2.

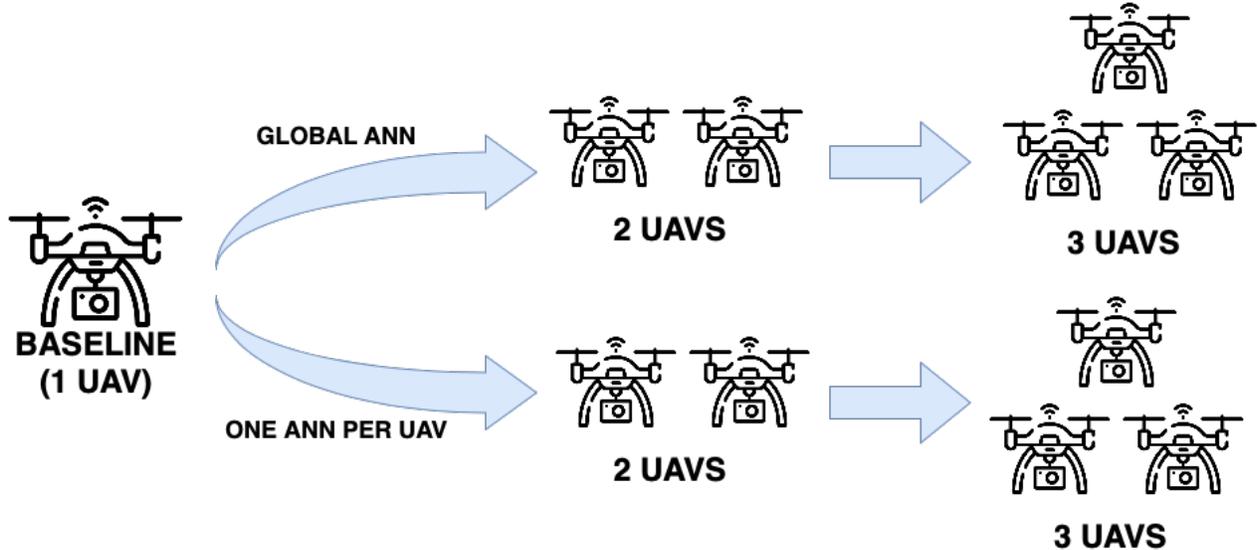

Figure 6: Diagram of how the experiments are conducted for each map. Initially, the system is tested with a given map and a single UAV. If a solution is obtained, the number of UAVs is increased as the system is able to solve the problem for that map. For the experiments, it is differentiated to use a single ANN for all UAVs and a global ANN.

### 4.2 Experimental Results

The results indicate that a global ANN is usually the best choice because it has the fastest solutions (Table 3). Although it is slower on 5x5 cell maps for 3 UAVs, it is only on the order of a few seconds. When using larger maps the model needs more time to find a solution regardless of the number of UAVs. This is due to the size of the exploration tree. That is, the larger the map, the more possible combinations of paths the ANNs have to evaluate. Usually, when using 3 UAVs these exploration times are slower compared to when using 2 UAVs. It is not a problem since it is often a few seconds and, the more UAVs, the easier it is to be able to resign the task of a fallen UAV to the others.

In all the experiments conducted the map has been completely covered at least once. The total number of episodes in an experiment where UAVs fly over all cells in the map is higher when a global ANN is used. This is indicative of the robustness of that ANN configuration to the problem. Having a greater number of solutions demonstrates that the system is learning and is able to find solutions as the randomness component $\epsilon$ is reduced.

Regarding the evolution of the flight paths (Figure 7), it was found to be highly dependent on the initialization and random component of all Q-Learning problems. That is to say, if in the first actions wrong decisions are made in the long execution it penalizes the solutions. In the example of Figure 7 the UAVs keep flying over the cells at the left of the map. In this case, the UAVs started the flight by taking many actions that resulted in passing through the cells on the left side. An excess of these actions caused the UAVs to barely fly over other cells.

Having solutions with different map sizes and different numbers of UAVs confirms that the system is generic enough to work under different conditions. From the authors' best knownledge, this is the only paper that can do that. Other papers only work with predefined map sizes [41, 42, 43].

### 4.3 Required Time Evolution

The speed to cover the entire map is also reflected in the time measure. Many episodes have a run time of 30 min. These coincide with the cases in which the entire map is not covered. However, in cases where this is not the case, the time required decreases as the training advances. It can be seen that it is highly dependent on the size of the map and the number of UAVs (Figure 8). In many cases, once the overall minimum is reached, the results worsen significantly. It is caused by the noise introduced by the random component of the chosen method. Because of this, the sequence of steps that is optimal is that of the episode of the sequence that finds the solution in the shortest time.

| Cell map size | ANN configuration | UAVs used | Solutions found | Minimum solution time |
|---|---|---|---|---|
| 5×5 | Baseline | 1 UAV | 28 out of 30 episodes | 00:00:40 |
| | One UAV per ANN | 2 UAVs | 26 out of 30 episodes | 00:02:09 |
| | | 3 UAVs | 27 out of 30 episodes | 00:01:10 |
| | Global ANN | 2 UAVs | 29 out of 30 episodes | 00:01:32 |
| | | 3 UAVs | 25 out of 30 episodes | 00:01:29 |
| 6×6 | Baseline | 1 UAV | 28 out of 30 episodes | 00:02:38 |
| | One UAV per ANN | 2 UAVs | 7 out of 30 episodes | 00:04:23 |
| | | 3 UAVs | 12 out of 30 episodes | 00:04:27 |
| | Global ANN | 2 UAVs | 22 out of 30 episodes | 00:02:12 |
| | | 3 UAVs | 14 out of 30 episodes | 00:04:39 |
| 7×7 | Baseline | 1 UAV | 24 out of 30 episodes | 00:03:14 |
| | One UAV per ANN | 2 UAVs | 9 out of 30 episodes | 00:06:17 |
| | | 3 UAVs | 12 out of 30 episodes | 00:07:01 |
| | Global ANN | 2 UAVs | 14 out of 30 episodes | 00:06:16 |
| | | 3 UAVs | 9 out of 30 episodes | 00:06:51 |
| 8×8 | Baseline | 1 UAV | 9 out of 30 episodes | 00:07:37 |
| | One UAV per ANN | 2 UAVs | 2 out of 30 episodes | 00:19:31 |
| | | 3 UAVs | 2 out of 30 episodes | 00:16:58 |
| | Global ANN | 2 UAVs | 5 out of 30 episodes | 00:14:14 |
| | | 3 UAVs | 5 out of 30 episodes | 00:13:52 |
| 9×9 | Baseline | 1 UAV | 8 out of 30 episodes | 00:12:17 |
| | One UAV per ANN | 2 UAVs | 1 out of 30 episodes | 00:24:45 |
| | | 3 UAVs | 1 out of 30 episodes | 00:20:53 |
| | Global ANN | 2 UAVs | 2 out of 30 episodes | 00:16:15 |
| | | 3 UAVs | 1 out of 30 episodes | 00:20:27 |

Table 3: Summary table with the observed results of the experiments. Results are displayed with the minimum time in each configuration needed for finding a solution. As in the papers discussed in Section 2, the results shown here are those obtained from a single execution due to the computational and time costs of averaging the results of multiple runs.

The evolution of the time taken by a fixed number of UAVs to find the solution on different maps is highly dependent on the size of the map. In Figure 9, an example plot with 2 UAVs is shown in which it can be seen that the time curve shows more growth than the curve of the number of cells in a map.

### 4.4 Taken Actions Evolution

The fraction of the actions where a new cell was discovered, also know as valid actions, among those taken shows different behavior. Therefore, it is a factor to be taken into account, as it also determines whether the system makes too many errors or takes too many cycles. The more UAVs, the more actions are performed, decreasing the overall percentage of valid actions among all actions taken (Figure 10). In the initial episodes, the UAVs take a lot of wrong actions. The accumulation of this amount of failures impairs the percentage of valid actions taken over the total number of actions. This is not a problem a priori, since the desired solution is reached faster in the next episodes.

The improvement of having more errors at the beginning comes from the fact that the map exploration tree is covered faster due to the simultaneous flight of the UAVs. As the exploration tree is covered faster, more information is extracted. Therefore, usually more UAVs in a swarm means that the task is performed faster in future episodes (Table 3). In the case it is slower, it may be an indicator that more episodes are needed in order to obtain a more optimal solution. The computational cost is very large considering that it is only a few seconds or minutes slower.

## 5 Discussion

Other papers described in Section 2 this paper proposes the use of an ANNs based o dense layers. This type of layer has also shown the ability to coordinate groups of UAVs. Moreover, being trained in each case with the memory of each UAV it seems to be able to assign correct actions to the UAVs without extracting spatial information from the map like convolutional networks. It may mean that the most important thing may not be the spatial relationship of the map, but the sequence of movements of each UAV without the noise of the actions taken by the other UAVs.

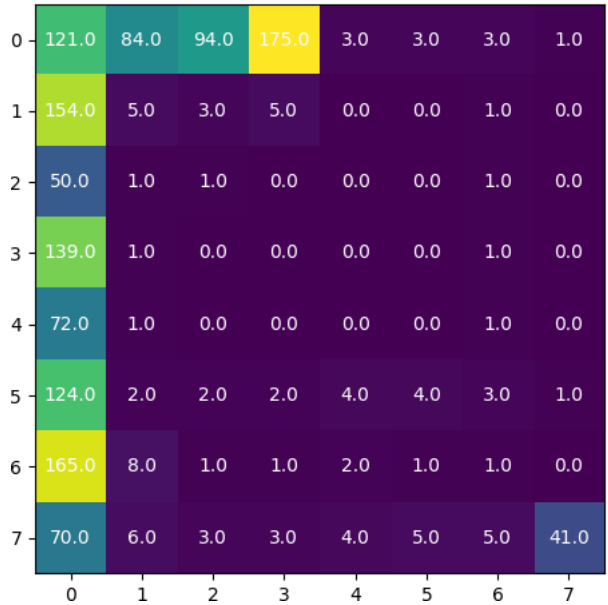

Figure 7: Example of a heatmap reflecting the number of times a UAV passes through each cell. In the first actions UAVs take from the beginning, the model took the wrong sequence of initial movements. This caused the UAVs to have a preference for traversing the left edge, thus consuming most of the time. In this way, many cells inside are left unvisited.

As in other papers in the state of the art, the system has been tested on squared cell maps. Unlike the other papers, it has been tested on different map sizes, not on fixed-size maps [44, 45]. The maps used do not present additional information, like those used in the other papers. That is, it is not necessary to add more information such as targets or distance maps. So it is not necessary to make previous studies of the map.

Since using a single global ANN for all UAVs usually requires less time than using one ANN per UAV indicates that the appropriate configuration is to use a global ANN. This means that paths calculated using a global network have fewer errors and loops, indicating that the paths are as direct as possible. The more direct they are, the shorter they are, determining that they are more optimal. In other multi-agent problems global ANNs were the best option like in the paper of Mnih et al. with their A3C algorithm [46].

The overall percentage of correct actions taken decreases with the increase of UAVs. This is due to the fact that in the first episodes too many wrong actions are taken because the agents do not have much knowledge. In spite of this, the solution is sometimes reached in less time due to the simultaneity of their flight and, the more UAVs, the more fault tolerance is ensured in case any UAV can't keep flying.

The time taken to explore the entire map is strongly affected by the number of cells on the map. The growth of the time taken exceeds the growth of the number of cells of the maps used in the experiments. This is mainly due to the size of the exploration space that ANN faces in order to find the best paths. The greater the number of cells, the larger the space. In addition, we have to add the sequences of the UAV paths, whether they are one or more. That is, a path has to be a correct sequence of adjacent and fully navigable cells, which adds complexity to the system by having to maintain this consistency as there are more and more cells. This is why the more cells, the more time the ANN requires obtaining correct paths.

## 6 Conclusions

In this work, a system capable of calculating the paths with the shortest flight time for UAV swarms using Q-Learning [26] techniques is proposed. To enhance the capabilities of these techniques, decision-making is done with the help of densely connected ANNs. Employing a single global ANN for all UAVs presents more solutions in less time. Finding models that find solutions quickly makes the system more portable to different systems. In this way, users will find it more convenient to use since money does not have to be spent on expensive systems. Typically, the cost savings can be invested in improving UAV features such as battery life by users.

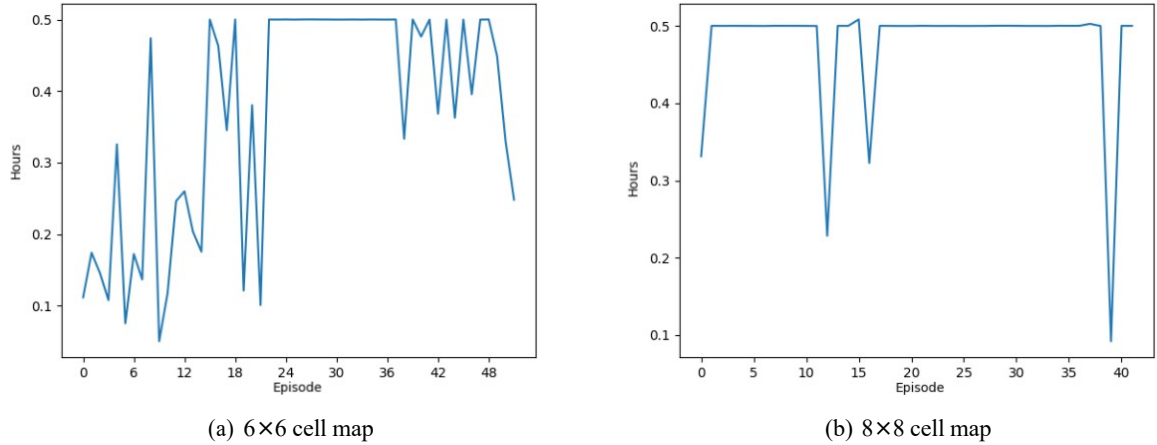

(a) 6×6 cell map      (b) 8×8 cell map

Figure 8: Example of two plots with the evolution of the hours (y-axis) consumed as the episodes elapse (x-axis) for different maps. In both, it can be seen that the solution with the shortest time can be followed by episodes with worse results. The optimal results are those episodes with the shortest duration. The shortest time implies that it is the one with the least number of incorrect actions and the least number of loops.

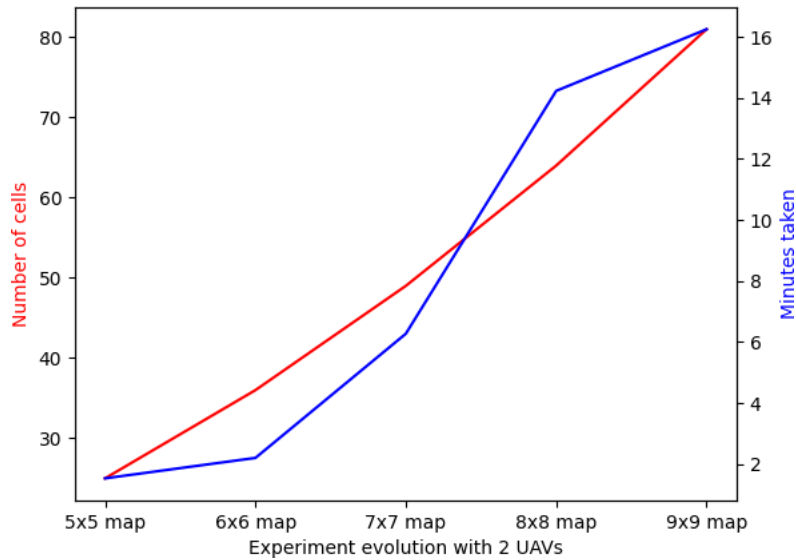

Figure 9: Comparison of the growth of the curve of the time taken for each map with respect to the growth of the curve of the number of cells contained in each map. The growth is greater in the time curve. Specifically, starting from the map of 8×8 cells.

The system is capable of obtaining satisfactory results with squared cell maps of different sizes. The evolution of the time required to find a solution as it increases faster than the increase in the number of cells in each experiment regardless of the number of UAVs in the swarm. Therefore, it is necessary to be able to adapt the size of the map to the activity to be carried out in order to get the best results as possible. Tasks that imply high altitude do not need as many cells because their sensors capture a large part of the terrain in each cell. Reducing the number of cells allows the system to make better and faster decisions due to the smaller exploration space.

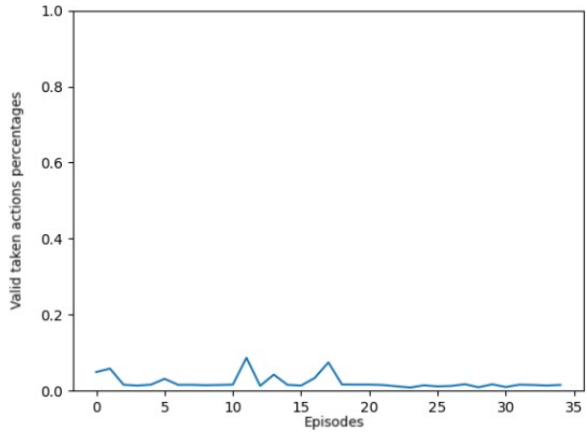

(a) 1 UAV

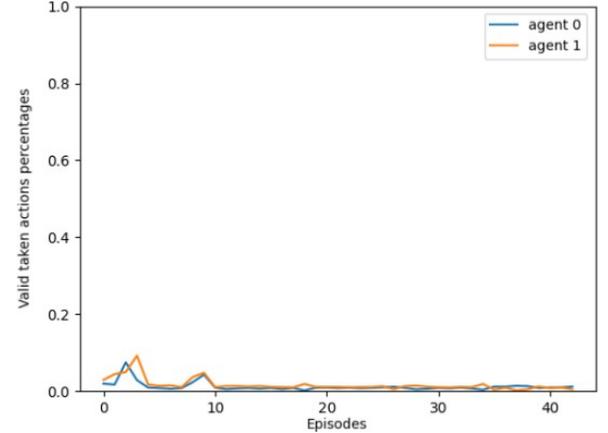

(b) 2 UAVs

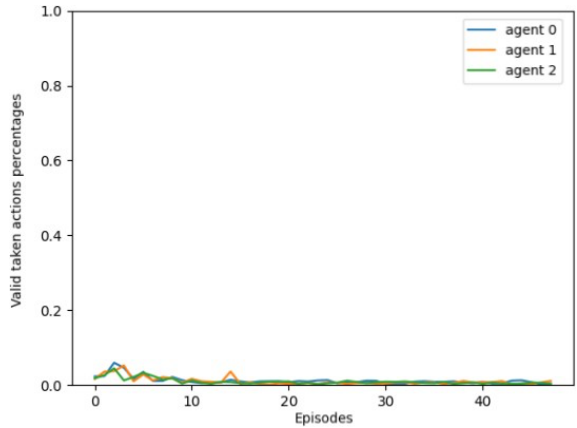

(c) 3 UAVs

Figure 10: Percentage of non-error actions (y-axis) taken by each UAV over the course of the episodes (x-axis). Each line of a different color symbolizes a UAV. The more UAVs the lower the percentage. This percentage is affected by the sum of the errors of all UAVs together. When taking the first actions UAVs make many mistakes. This accumulation of errors hurts the percentage of valid actions taken by the UAVs as a whole.

It is not necessary to provide additional information to the map to direct the paths. Therefore, the system is able to calculate the optimal paths having only the information of the cell map, the current position of the UAVs on the map and the evolution of the flight paths along the map. Using so little information avoids having to know the terrain in advance. If information is to be added to guide the UAV paths, it is necessary to make such a study. Therefore, many users may end up discarding the use of the system due to this added complexity. On the other hand, if it is necessary to guide the paths, the system can be biased because user errors can be made that prevent better paths from being found.

## 7 Future Work

From the starting point of this paper, some bases are laid for the creation of other systems capable of working on different maps. In this way, generic systems with commercial potential can be obtained.

In the future, efforts will be made to improve these results with a previous study about the optimal initial distribution of the UAVs on the map should be carried out. Also, models will be trained on cell maps optimally divided according to the resolution of the UAV cameras.

Future developments will include experiments with 3D maps in which more movements such as pitch and roll will be possible.

Experiments will be made with maps with obstacles in order to make agents learn how to reduce the risks during the flight. Obstacles can be fixed obstacles (trees, poles...) or dynamic obstacles (birds, other UAVs...).

### Acknowledgments

This work is supported by Instituto de Salud Carlos III, grant number PI17/01826 (Collaborative paper in Genomic Data Integration (CICLOGEN) funded by the Instituto de Salud Carlos III from the Spanish National plan for Scientific and Technical Research and Innovation 2013–2016 and the European Regional Development Funds (FEDER)—"A way to build Europe.". This paper was also supported by the General Directorate of Culture, Education and University Management of Xunta de Galicia ED431D 2017/16 and "Drug Discovery Galician Network" Ref. ED431G/01 and the "Galician Network for Colorectal Cancer Research" (Ref. ED431D 2017/23). This work was also funded by the grant for the consolidation and structuring of competitive research units (ED431C 2018/49) from the General Directorate of Culture, Education and University Management of Xunta de Galicia, and the CYTED network (PCI2018_093284) funded by the Spanish Ministry of Ministry of Innovation and Science. This paper was also supported by the General Directorate of Culture, Education and University Management of Xunta de Galicia "PRACTICUM DIRECT" Ref. IN845D-2020/03. The experiments described in this section were carried out using the equipment of the Galician Supercomputing Center (CESGA).

### Supplementary Material

Source code is available at: `https://github.com/TheMVS/uav_swarm_reinforcement_learning`

A Docker container is available at: `https://hub.docker.com/r/themvs/uav_swarm_reinforcement_learning`